\documentclass[showpacs,superscriptaddress,twocolumn,aps,amsmath,amssymb,prb]{revtex4}
\usepackage{amsfonts}
\usepackage{graphicx}

\newcommand{\cf}{cf.\ }
\newcommand{\be}{\begin{equation}}
\newcommand{\ee}{\end{equation}}
\newcommand{\bea}{\begin{eqnarray}}
\newcommand{\eea}{\end{eqnarray}}
\newcommand{\Fig}[1]{Fig.\,\ref{#1}}
\newcommand{\Eq}[1]{Eq.\,(\ref{#1})}

\newcommand{\nl}{\nonumber \\}

\begin{document}
\title{Peierls Instability Induced Ferromagnetic Insulator
at Orbital Order Transition}
\author{J.~H.~Wei}
\email{wjh@ruc.edu.cn}\affiliation{Department of Physics, Renmin
University of China, Beijing 100872, China}
\author{D.~Hou}
\affiliation{Department of Physics, Renmin University of China,
Beijing 100872, China}\affiliation{Department of Physics, Shandong
University, Jinan 250100, China}
\author{X.~R.~Wang}
\affiliation{Department of Physics, Hong Kong University of
Science and Technology, Kowloon, Hong Kong}
\date{\today}

\begin{abstract}
The origin of ferromagnetic insulating state of La$_{7/8}$Sr$
_{1/8}$MnO$_3$ is investigated. Based on the tight-binding model, it
is shown that this state can be attributed to the Peierls
instability arisen from the interplay of spin and orbital ordering.
The importance of the hole-orbiton-phonon intercoupling in doped
manganites is revealed. This picture explains well the recent
experimental finding of the reentrance of ferromagnetic metal state
at low temperature [Phys. Rev. Lett. 96, 097201 (2006)].
\end{abstract}

\maketitle

\section{INTRODUCTION}%
R$_{1-x}$A$_x$MnO$_3$ (R being rare-earth ions and A for divalent
ions, e.g. A$=$ Ca, Sr, Ba or Pb) have been intensively studied for
more than a decade because of its rich physics, such as unusual
colossal magnetoresistance (CMR)\cite{Sch95}, interesting
ferromagnetic (FM) and antiferromagnetic (AFM) phases, and charge
and/or orbital ordering. Despite intensive research and progress
made so far in understanding the system, the origin of the
ferromagnetic insulating (FMI) phase observed in lightly doped
La$_{7/8}$Sr$_{1/8}$MnO$_3$ \cite{Uru95} is still controversial and
unknown. CMR can be qualitatively understood in terms of the
celebrated double exchange (DE) model \cite{And55}, which means the
simultaneous appearance of ferromagnetism and metallic behavior.
Thus, FMI is not compatible with  the DE model, at least not in a
direct way. Although it is generally recognized that the FMI state
may relate to the interplay of charge, spin and orbital degrees of
freedom, no concrete picture is available. Finding such a picture is
the focus of this paper.

It shall be useful to first summarize the experimental observed
phases of La$_{7/8}$Sr$_{1/8}$MnO$_3$ solid. Lowering the
temperature, the first structural phase transition from orthorhombic
to monoclinic orders occurs at $T_{\rm JT}\sim283$K. The system is
in the orthorhombic paramagnetic insulating phase (phase I) above
$T_{\rm JT}$, and monoclinic paramagnetic insulating phase (phase
II) below $T_{\rm JT}$ but above ferromagnetic (FM) transition
temperature $T_{\rm C}\sim183$K. The second structural phase
transition from monoclinic to triclinic orders occurs at $T_{\rm
OO}\sim150$K, separating the monoclinic ferromagnetic metallic phase
\cite{Cox01,Gec04} (phase III) from the triclinic FMI state (phase
IV) having a superstructure with unit cell $2a_c \times 4b_c\times
4c_c$ ($a_c\times b_c\times c_c$ is the unit cell of the
high-temperature orthorhombic phase) \cite{Tsu01}. The FMI state of
phase IV is supported by thermal activated exponential
$T$-dependence \cite{Uru95,Liu01} of the resistivity that increases
rapidly with the decrease of temperature for $T<T_{\rm OO}$, and the
metallic nature of phase III is confirmed by the fact that the
resistivity decreases with temperature below $T_{\rm C}$
\cite{Uru95}. The system undergoes another metal-insulator
transition from FM insulator to FM metal (V) at temperature
$T_R\simeq 30K$. Phase V is a reentrance of FM metallic (FMM) state
through the FMI phase in $T_{\rm R}<T<T_{\rm OO}$, newly discovered
by nuclear magnetic resonance technique \cite{Pap06}. The crystal
structure below $T_{\rm R}$ is still unknown. Table 1 lists all five
phases, characterized by four critical temperatures. So far, phases
I-III have been widely studied and well understood in terms of
crystal structure, magentic and transport properties.
\begin{table}[htbp]
\caption{Phase transitions of La$_{7/8}$Sr$_{1/8}$MnO$_3$:
P$-$Paramagnetic, FM$-$Ferromagnetic, I$-$Insulator and
M$-$Metallic} \label{tab1} \centering
\begin{tabular}{|c|c|c|c|c|}
  \hline
  Phase & Temperature & Structure & Magnetism & Transport \\
  \hline
  I & $T>T_{\rm JT}$ & Orthorhombic & P & I \\
  II & $T_{\rm C}<T<T_{\rm JT}$ & Monoclinic & P & I \\
  III & $T_{\rm OO}<T<T_{\rm C}$ & Monoclinic& FM & M \\
  IV & $T_{\rm R}<T<T_{\rm OO}$ & Triclinic & FM & I \\
  V & $T<T_{\rm R}$ & Unknown & FM & M \\
  \hline
\end{tabular}
\end{table}

In order to explain the FMI state of La$_{7/8}$Sr$_{1/8}$MnO$_3$,
many different models were proposed in the literature, including
charge polarons order \cite{Yam96}, checkerboard-like charge order
\cite{Yam00}, orbital order without charge order \cite{End99}, etc.
Most of them are not consistent with the recent resonant X-ray
scattering (RXS) experiments, which observed an orbital polaron
lattice (OPL) in phase IV \cite{Gec05}. This observation is the
basis of FMI$-$orbital polaron model, proposed by Kilian and
Khaliullin \cite{Kil99}. Orbital polarons can be viewed as charge
carriers (holes) dressed by the inter-connected orbital states. At
light doping magnanites, one Mn$^{4+}$ site is surrounded by six
neighboring Mn$^{3+}$ sites, equivalent to one hole (with $e_g$
orbits unoccupied) surrounded by six (occupied) $e_g$ orbits. The
strong hole-orbital coupling will polarize those $e_g$ orbits and
make them point towards the hole to minimize the interaction energy
as well as the kinetic energy \cite{Kil99}. The formation of OPL in
La$_{7/8}$Sr$_{1/8}$MnO$_3$ emphases the importance of hole-orbital
interaction for the FMI state \cite{Gec05,Kil99,Miz00}.

Obviously the genuine mechanism of the FMI state should explain
following experimental features:  1) the measured $\rho-T$ curves
\cite{Uru95,Liu01}; 2) the giant phonon softening \cite{Cho05}; and
3) the reentrance of the FMM state (IV$\rightarrow$V) \cite{Pap06}.
It is difficult for orbital polaron model to explain feature 1) and
3). The reasons are as follows. Firstly, the orbital polaron cannot
reproduce the measured $\rho-T$ curves of La$_{1-x}$Sr$_x$MnO$_3$ at
$T<T_{\rm OO}$ (\cf~\Fig{fig3}). Polaron hopping model yields
$\rho=\rho_0\exp(E_a/k_BT)$, where $E_a$ is the activation energy of
polarons, from which the binding energy ($E_b=2E_a$) of a polaron
can be deduced. The fitting to experimental data \cite{Liu01,Gec05}
with hole-polaron hopping model gives $E_b\sim 0.06-0.12$eV, about
an order of magnitude smaller than the theoretical value ($E_b\sim
0.6$ eV in Ref.\onlinecite{Kil99}). Secondly, the observed
reentrance of the FMM state at lower temperature is inconsistent
with polaron model since polaron hopping becomes more and more
difficult when the temperature is lowered. Therefore, the orbital
polaron itself is not sufficient to explain the FMI state and the
origin of the metal-insulator transitions at $T_{\rm OO}$ remains
intriguing and controversial.

The above difficulties can be removed by including electron-phonon
({\it e-ph}) interaction in the simple electron-orbit coupling
model. We argue that the mechanism of the FMI state is the {\it
e-ph} coupling induced Peierls instability \cite{Pei55} that opens
an energy gap at $T<T_{\rm OO}$. In this paper, we shall show that
the Peierls instability can explain the observed reentrance of FMM
state quantitatively, and the other experimental features at the
qualitative level.

\section{THEORY}%

\begin{figure}
\includegraphics[width=3.3in]{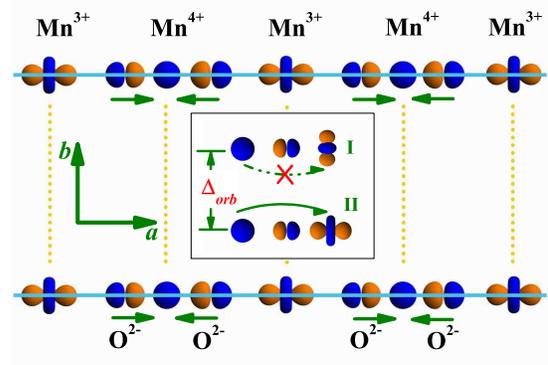}
\caption{(color online) The orbital order induced quasi-1D
confinement of the motion of holes in $a$ direction. The dashed
lines denote the weak interchain couplings in $b$ and $c$
directions. The arrows near the O$^{2-}$ ions indicate the
displacement of the oxygen sublattice. The insert shows the orbital
order induced effective coupling between adjacent Mn$^{3+}$ and
Mn$^{4+}$ sites mediated by the oxygen $2p$ orbit. Configuration I
and II correspond to different occupation of $e_g$ orbitals on the
Mn$^{3+}$ site. $\Delta_{orb}$ is the splitting of energy between I
and II \cite{Kil99}. \label{fig1}}
\end{figure}

Our theoretical model is based on the quasi-one-dimensional (1D)
confinement of the motion of holes in La$_{7/8}$Sr$_{1/8}$MnO$_3$,
which has been demonstrated by following experiments: 1) the
convergent-beam electron diffraction and selected-area electron
diffraction seeing the superstructure ($2a_c\times4b_c\times4c_c$)
of the FMI phase \cite{Tsu01} (anisotropic 3D motion); 2) the
resonant X-ray scattering showing an alternation of hole-rich and
hole-poor planes in $c$ direction \cite{Gec05}, which confines holes
to move in the two-dimensional hole-rich $a-b$ planes (2D motion);
and 3) the resonant X-ray scattering also revealing the formation of
OPL \cite{Gec05}, which further confines holes to move along the
one-dimensional charge stripes in $a$ direction (1D motion, see
\Fig{fig1} and the following explanations). It is known that the
quasi-1D confinement originates from the orbital order and its
induced effective coupling between adjacent Mn$^{3+}$ and Mn$^{4+}$
sites. The insert of \Fig{fig1} depicts two possible configurations
of one Mn$^{4+}-$O$^{2-}-$Mn$^{3+}$ unit with different $e_g$
orbital occupations on the Mn$^{3+}$ site. Configuration II
corresponds to $e_g$ orbit pointing towards the hole along the axis
of oxgen $2p$ orbit while configuration I corresponds to $e_g$ orbit
pointing towards other directions. Due to orbital anisotropy,
configuration II results in a maximized overlapping of wavefunctions
of occupied $e_g$ and $2p$ orbits. That induces two direct effects:
1) lower energy of configuration II than that of I with an energy
difference $\Delta_{orb}$ \cite{Kil99}; 2) effective
Mn$^{3+}-$Mn$^{4+}$ coupling with considerable transfer integration
($t\neq0$). As a result, configuration II is stable and favors the
motion of holes along $a$-axis. In the hole-rich planes shown in
\Fig{fig1}, configuration II periodically repeats itself along
$a$-direction thus forms the quasi-1D
Mn$^{3+}-$O$^{2-}-$Mn$^{4+}-$O$^{2-}$ chain-like pathways of hole
transport. Please be noted that weak interchain couplings
($t_\perp$) in $b$ and $c$ directions also exist, resulting from the
hole transfer between unoccupied and occupied $e_g$ orbits
\cite{Gec05}. Why the orbital order induced quasi-1D confinement can
only be observed around $x=1/8$? Qualitatively speaking, the orbital
disorder-order transition takes place at light hole doping provided
that the Jahn-Teller phonons and superexchange processes mediate an
effective coupling between orbits on neighboring sites \cite{Kil99}.

Now we are ready to present our model Hamiltonian to describe the
quasi-1D motion of particles (electrons/holes). As shown in
\Fig{fig1}, the particles moving along
Mn$^{3+}-$O$^{2-}-$Mn$^{4+}-$O$^{2-}$ chains will simultaneously
couple to the orbits on the manganese sites and lattice
displacements (phonons) on the oxygen ones. Considering that the
quasi-particles of orbital degree of freedom are bosonic orbitons
\cite{Ish96, Sai01}, we can establish an electron-orbiton-phonon
intercoupling model. For simplicity, only Mn sites are included in
the model and the oxygen degrees of freedoms are integrated out,
giving rise to the effective Mn-Mn coupling and the phonon
modulation to it [cf~\Eq{Ham}]. Then, the tight-binding model 
can be written as ($\hbar=1$ and $k_B=1$ are assumed, and they  
are restored in final results),
\bea\label{Ham} H&=&t\sum_{j}(c_j^+c_{j+1}+h.c.)+\sum_q\epsilon_qa_
q^+a_q+ \sum_{j}c_j^+c_j\mathcal{Q}_{j}\nl
&+&\sum_p\omega_pb_p^+b_p+\sum_{j}(c_{j}^+c_{j+1}\mathcal{P}_j+h.c.)\nl
\mathcal{Q}_{j}&=&(Q_{j-1}+Q_{j+1})\nl
\mathcal{P}_{j}&=&(P_{j+1}-P_{j}) \eea
with
\bea\label{NQP} Q_j&=&\sum_qR_qe^{iqR_j}(a_q+a_{-q}^+)\nl
P_j&=&\sum_pG_pe^{ipR_j}(b_p+b_{-p}^+)\eea
where $c_j^+$ is the creation operator of a particle on site $j$
($R_j$ is the atom position), $a_q^+$ ($b_p^+$) the creation
operators of a orbiton (phonon) with momentum $q$ ($p$),
$\epsilon_q$ and $\omega_p$ are the dispersion spectrum of orbiton
and phonon respectively. The third term is the {\it e-orb} coupling
between a particle on site $j$ and the orbitons on the nearest
neighbor sites $j\pm1$, and the last term describes the {\it e-ph}
coupling due to the phonon modulation of the particle hopping
between sites $j$ and $j+1$ \cite{Man00} (see \Fig{fig1}). $R_q$ and
$G_p$ are the {\it e-orb} and {\it e-ph} coupling constant
respectively, which satisfy $G^*_{-p}=G_p$ and $R^*_{-q}=R_q$ to
ensure the Hermitian character of \Eq{Ham}.

What are the different roles of the {\it e-orb} and {\it e-ph}
interactions in \Eq{Ham}? One can see that the orbiton couples to
the on-site charge density $n_j$, which promotes the formation of
Holstein-like small polarons in the strong-coupling limit
\cite{Hol59}. That is the physical origin of the formation of
orbital polarons \cite{Kil99, Gec05}. The essential effect of
Holstein-like small polarons is reducing the electron bandwidth (or
enlarging the electron effective-mass). This interaction predicts a
transition from band transport to thermal-activated hopping of
polarons at high temperature, but it can not open an energy gap to
induce the metal-insulator transition at low temperature. However,
the {\it e-ph} interaction in \Eq{Ham} can take the role of opening
the energy gap, since the phonon modulation of the electron transfer
predicts the Peierls instability in quasi-1D systems at low
temperature \cite{Pei55, Su80}. For La$_{7/8}$Sr$_{1/8}$MnO$_3$,
when the orbital order effectively confines the particles to move
along quasi-1D pathways (see \Fig{fig1}), the strong {\it e-ph}
interaction distorts the oxygen sublattice to make O$^{2-}$ ions
displace towards Mn$^{4+}$, which enhances the polarization of
orbits and helps to stabilize the orbital order (see \Fig{fig1}).
There thus forms the quasi-1D charge-density-wave (CDW) state in
manganese sublattice accompanied with bond-order-wave (BOW) state in
oxygen sublattice with twice the period of the original lattices
(dimerization). One analogous material with similar CDW/BOW ground
state is the halogen-bridged mixed-valence metal complex which has
been well studied with this kind of {\it e-ph} interaction in the
literature \cite{Gam92}.

At $T<T_{\rm C}$, the ferromagnetic $t_{2g}$ spins of Mn ions
exclude the opposite spin to occupied $e_g$ orbitals due to the
Hund's rules, i.e. only carriers with the majority-spin contributes
to the charge transport. Before coupling to orbitals and phonons,
the Mn$^{4+}-$Mn$^{3+}$ transfer [the first term in \Eq{Ham}] in
majority-spin subspace conduces a half-filled metallic energy band
and the bare particle spectrum $\xi_k^0$ can be written as
\be\label{EK0} \xi_k^0=\gamma_k^0+\delta_k^0 \ee
with
\bea\label{ESP0} \gamma_k^0&=&-2t\cos k_xa \nl
\delta_k^0&=&-2t\eta(\cos k_yb+\cos k_zc) \eea
where the effect of weak interchain coupling ($t_\perp=\eta t$,
$\eta\ll 1$) has been included in $\delta_k^0$. $\eta$ is a
dimensionless parameter denoting relative strength of the interchain
coupling to intrachain one.

The orbitons and phonons then couple to the tight-binding particles,
as described in \Eq{Ham}, and change above metallic band to a
insulator one. The simultaneous couplings between two bosons to one
fermion are technically difficult to solve, even though the direct
orbiton-phonon coupling has been ignored in \Eq{Ham}. Thus we make
two approximations for {\it e-orb} coupling in following
calculations. The first one is the $q$-independence $\epsilon_q
(=\epsilon)$ and $R_q(=r)$, based on the theoretical result that the
orbiton is almost dispersionless at strong {\it e-orb} couplings
\cite{Bal02}. The second one is the perturbation theory of small
polarons \cite{Man00} at strong {\it e-orb} couplings, which
conduces
\be\xi_k=\xi_k^0\exp(-r^2/\epsilon^2)-\frac{r^2}{\epsilon} \equiv
A\xi_k^0-B\ee
where $A$ denotes the electron bandwidth reduction (or
effective-mass enhancement) and $B$ the energy shift due to strong
{\it e-orb} couplings. The above approximations have captured the
essential physics of {\it e-orb} coupling: electron effective-mass
enhancement ($A$) and energy shift ($B$), so that it should produce
reasonable results in the limit of strong {\it e-orb} interactions.
Please be noted that above approximations are only valid at finite
$\epsilon$ to avoid $B$ unphysically large. After electron coupled
to orbital degree of freedom,
\be\label{EK1} \xi_k=\gamma_k+\delta_k \ee
where
\bea\label{ESP1} \gamma_k&=&A\gamma_k^0 \nl
\delta_k&=&A\delta_k^0-B
\eea

Let us make some further comments on the {\it e-orb} interactions concerned
in the present work. Our model Hamiltonian describes electrons
coupled to two bosonic fields (orbitons and phonons), which is formally
similar to the model of electrons coupled to two kinds of phonons for
cuprates \cite{Yon92}. The key feature of orbitons distinguishing
them from usual phonons is the strong spatial anisotropy due to the
symmetry of $e_g$ electron wave functions, which has been
qualitatively considered in building up our quasi-1D Hamiltonian
[see \Fig{fig1} and \Eq{Ham}]. The quantitative description of
orbitons by means of pseudospin operators, which may be required for more
general study \cite{Kil99}, has been beyond the scope of this
paper. The parameter characterizing the orbital ordering come into
play due to orbitons is $\lambda_o\equiv r/t$. In our calculations,
$\lambda_o\sim 0.2-0.65$, which is realistic to manganites and
consistent with literatures \cite{Kil99}.

Particles with new energy spectrum in \Eq{EK1} then couple to phonon
degree of freedom. Now, we can analytically solve the Hamiltonian by
transforming it into momentum space,
\be\label{Hamk}
H=\sum_k\xi_kc_k^+c_k+\sum_p\omega_pb_p^+b_p+\sum_{k,p}g_{pk}c_{k+p}^+c_k(b_p+b_{-p}^+)
\ee
where $g_{pk}$ is the {\it e-ph} coupling constant in momentum space
and its dependence on $k$ is to be neglected in what follows
($g_p=g^*_{-p}\propto M_p$), as in the literature \cite{Sad06}. That
Hamiltonian for 1D systems indicates a Peierls superstructure which,
described by introduction of the following anomalous average
\cite{Bog91}, breaks translational symmetry of initial lattice:
\be\label{AVE}\Delta=g_{2k_F}<b_{2k_F}+b_{-2k_F}^+>\neq0 \ee
where angular brackets denote thermodynamic average that can be
obtained by performing Gibbs average to Matsubara equations of
motion for operators $b_K$ and $b_{-K}^+$ ($K=2k_F$ for Peierls
phase transition) \cite{Sad06},
\bea\label{EOM}
(-\frac{\partial}{\partial\tau}-\omega_K)<b_K(\tau)>&=&\sum_kF(k,\tau=-0)
\nl
(-\frac{\partial}{\partial\tau}+\omega_K)<b_{-K}^+(\tau)>&=&-\sum_kF(k,\tau=-0)
\eea
where $F(kt)=-i<Ta_k(t)a_{k-K}^+(0)>$ is the anomalous Green's
function describing the elementary {\it Umklapp} scattering process
$k-K\rightarrow k$.

After Fourier transformation over Matsubara ``time" to \Eq{EOM}, we
have:
\be
<b_K+b_{-K}^+>_{\omega_m}=-\frac{2g_K\omega_KT}{\omega_m^2+\omega_K^2}%
\sum_{k,n}F(k,\varepsilon_n)\ee
The condition for the Peierls phase transition is $\omega_m=0$ at
$K=2k_F$, i.e.,
\be\label{DLT}\Delta=g_K<b_K+b_{-K}^+>_{\omega_m=0}=-\frac{2g_K^2T}{\omega_K}%
\sum_{k,n}F(k,\varepsilon_n) \ee
In coordinate representation \Eq{AVE} describes Peierls deformation
potential characterized by the wave vector $K$: $V(x)=\Delta
e^{iKx}+\Delta^*e^{-iKx}$. The anomalous Green's function $F$ can be
derived from the Gorkov equations for Matsubara Green's functions
(limit to first order in $V$) under the ``nesting condition":
$\gamma_{k-K}=\gamma_{k-2k_F}=-\gamma_k$ \cite{Sad06},
\bea\label{Mat}
   G(k,\varepsilon_n) &=&
   G_0(k,\varepsilon_n)+G_0(k,\varepsilon_n)\Delta F(k,\varepsilon_n)
\nl F(k,\varepsilon_n) &=&
   G_0(k-K,\varepsilon_n)\Delta^*G(k,\varepsilon_n), \eea
which gives the following solutions,
\bea\label{FKE} G(k,\varepsilon_n)&=&\frac{i\varepsilon_n-\delta_k+
\gamma_k}{(i\varepsilon_n-\delta_k)^2-\gamma_k^2-\Delta^2}
\nl F(k,\varepsilon_n)&=&
\frac{\Delta^*}{(i\varepsilon_n-\delta_k)^2-\gamma_k^2-\Delta^2}.\eea
By replacing $i\varepsilon_n$ with $\varepsilon$, the new energy
spectrum is determined by the zero of denominators (pole) of
\Eq{FKE}:
\be\label{NESP}\varepsilon=\delta_k\pm\sqrt{\gamma_k^2+\Delta^2}.\ee
Inserting \Eq{ESP1} into \Eq{NESP} gives the energy gap $E_g$ of the
system,
\be\label{EG}E_{\rm g}=2\Delta-8A\eta t.\ee

By inserting \Eq{FKE} into \Eq{DLT} and then performing standard
calculations, we obtain the self-consistent equation of $\Delta$
(set to be real),
\be\label{CriT}
1=\frac{g^2}{2}\sum_k\frac{\sinh\frac{\sqrt{\gamma_k^2+\Delta^2}}{T}}%
{\cosh\frac{\sqrt{\gamma_k^2+\Delta^2}}{T}+\cosh\frac{\delta_k}{T}}%
\frac{1}{\sqrt{\gamma_k^2+\Delta^2}}.\ee
where $g=g_K\sqrt{2/\omega_K}$.

\section{DISCUSSION}%
So far we have derived the insulating state at $T<T_{\rm C}$
resulting from the Peierls instability at the orbital order
transition temperature $T_{\rm OO}$. With this mechanism, we first
quantitatively explain the observed reentrance of FMM state. The key
point is that a $T$-dependent interchain coupling $\eta(T)$ may
induce two critical temperatures $T_{\rm c1}$ and $T_{\rm c2}$ of
the metal-insulator transition. The former corresponds to the
ordinary Peierls instability and the latter to the reentrance of the
metallic state. It is called stepped Peierls transition theory, with
which Zhou and Gong explained the anomalous transport property in
NbSe$_3$ \cite{Zho88}. With this theory, we can elucidate the
reentrance of FMM state in La$_{7/8}$Sr$_{1/8}$MnO$_3$. For strong
{\it e-orb} coupling, the self-consistent equation of $T_{\rm MI}$
[derived from $E_g(T_{\rm MI})=0$] can be expressed as
\be\label{TMI} \Delta(T_{\rm MI})=4A\eta(T_{\rm MI}) t. \ee

In principle, $\eta(T)$ should increase with the decrease of
temperature, in analogy with the role of pressure. Since the exact
expression of $\eta(T)$ is hard to determined, we take the leading
linear term of its series expansion at low
temperature ($T<T_{\Delta}$),
\be\label{ETAT}\eta(T)=\eta_0(1-\alpha T/T_{\Delta}),\ee
where $T_{\Delta}$ denotes the critical temperature of the
charge order [$\Delta(T_{\Delta})=0$], which coincides with $T_{\rm
c1}$ at $\eta=0$ [\cf~\Eq{EG}] but is a little higher than $T_{\rm
c1}$ at $\eta>0$. $\eta_0$ is the $T$-independent part of $\eta$ and $\alpha$ 
a phenomenological parameter. In our calculations, we fix $T_{\Delta}=180$K by
considering the experimental $T_{\rm c1}=T_{\rm OO}\sim 150$K.
Please be noted that at least two parameters in the self-consistent
equation [\Eq{CriT}] dependents on each other after $T_{\Delta}$ is fixed.

\begin{figure}
\includegraphics[width=3.3in]{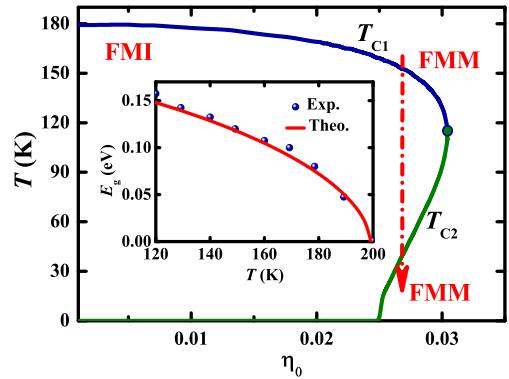}
\caption{(color online) Two critical temperatures of metal-insulator
transitions ($T_{\rm c1}$ and $T_{\rm c2}$) as a function of
interchain coupling strength $\eta_0$. The
large dots are the coincident points of $T_{\rm c1}$ and $T_{\rm
c2}$. The reentrance of the ferromagnetic metal state is shown by
the big arrow. The insert shows the temperature dependence of the
energy gap. The line is the theoretical result while the dots are
experimental data from Ref.~\onlinecite{Che08}. See text for details
of the parameters.}
\label{fig2}
\end{figure}

We then self-consistently solve \Eq{CriT}--(\ref{ETAT}) and
illustrate the change of $T_{\rm c1}$ and $T_{\rm
c2}$ with $\eta_0$ in \Fig{fig2}. The parameters are chosen as $r=0.2t$,
$g=1.1\sqrt{t}$, $\epsilon=0.6t$, $\alpha=0.5$ and $t=0.4$ eV. As shown in \Fig{fig2},
our calculations can yield $T_{\rm c1} \sim 150$K and
$T_{\rm c2}\sim30$K at $\eta_0\sim 0.028$, comparing well with the
experimental observation in Ref.~\onlinecite{Pap06} as indicated by
the big arrows in the figure. No reentrance of metallic state occurs
at small interchain coupling ($\eta_0<0.024$), characterized by
$T_{\rm c2}=0$. $T_{\rm c1}$ decreases slightly with increasing of
$\eta_0$ in this limit. Nonzero $T_{\rm c2}$ emerges at about
$\eta_0\sim 0.024$. Further increasing interchain coupling will
induce rapid increasing $T_{\rm c2}$ and rapid decreasing $T_{\rm
c1}$ meanwhile. When $T_{\rm c1}$ and $T_{\rm c2}$ coincide into one
point at large interchain coupling, the Peierls phase transition
disappears. This is just the interchain-coupling-induced
delocalization of quasi-1D states, as occurs in conjugated polymers
\cite{Mar06}. Since the value of $t$ for $e_g$ electron system is
believe to be $0.3\sim 0.6$ eV \cite {Kil99,Ede07}, our
theory quantitatively explains the reentrance of the FMM state
with consistent parameters in the literature.

In a recent experiment on the (011)-oriented La$_{7/8}$Sr$_{1/8}$MnO$_3$ films,
Chen {\it et al.} measured an energy gap opening at $T\sim190$K and increasing to
$\sim 0.16$ eV at $T=120$K \cite{Che08}. The absent physics mechanism
in their experiment has been already elucidated in this paper, that is, the observed energy
gap results from the Peierls instability at orbital order transition. To further clarify this point, we calculate the temperature dependence of the energy gap self-consistently, and show the result in the insert of \Fig{fig2} together with the experimental data. Some of above
parameters are adjusted to fit Chen {\it et al.}'s experiment: $T_{\Delta}=200$K,
$r=0.64t$ and $\eta_0=0.02$. As shown in the figure, our calculated $E_{\rm g}-T$ curve
is in good agreement with the experimental data.

\begin{figure}
\includegraphics[width=3.4in]{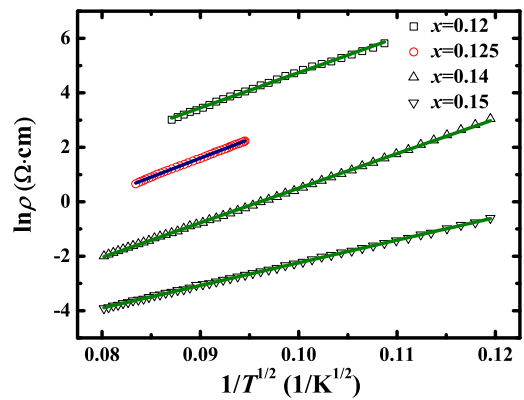}
\caption{(color online) The fits of the experimental measurements of
$\rho$ by $\rho=\rho_0\exp(T_0/T)^{1/2}$. The experimental data is
from Ref.~\onlinecite{Gec05} ($x=0.125$) and Ref.~\onlinecite{Liu01}
(for others).} \label{fig3}
\end{figure}

We then systematically study the effect of the {\it e-orb} coupling on
the Peierls phase transition, and find that large $r$ increases
$T_{\rm c1}$ and decreases, in meanwhile, $T_{\rm c2}$. As a
consequence, strong {\it e-orb} coupling will enlarge the phase area
of the FMI state in \Fig{fig2}. Sufficient strong {\it e-orb}
coupling will annihilate the reentrance of the FMM state by
decreasing $T_{\rm c2}$ to zero (the figure not shown). That
confirms the orbital order being in favor of the Peierls insulating
phase. Please be noted that the Peierls instability can not take
place without the orbital order induced quasi-1D confinement in
La$_{7/8}$Sr$_{1/8}$MnO$_3$.

With our theory, we can also explain the other experimental features
listed in the introduction at the qualitative level. The Peierls
instability opens an energy gap [see \Eq{EG}] and makes
La$_{7/8}$Sr$_{1/8}$MnO$_3$ undergo a metal-to-insulator transition
at $T_{\rm OO}$. As a consequence, the charge carriers transport
only through the localized gap states at $T<T_{\rm OO}$, which will
trigger the variable range hopping (VRH) \cite{Shk84}. That picture
is confirmed by the the good fit of experimental $\rho-T$ curve
\cite{Gec05,Liu01} with VRH mechanism: $\rho=\rho_0\exp(\sqrt
{T_0/T})$ in \Fig{fig3}. The exponent $1/2$ results from a parabola
density of states (DOS) with zero-DOS at the Fermi level $E_{\rm F}$
(called Coulomb gap). As regards the the giant phonon softening of
the Mn-O breathing mode \cite{Cho05}, we comment that it results
from the celebrated giant Kohn anomaly at Peierls instability, i.e.
the suppression of phonon frequency at $p\sim 2k_F$ \cite{Koh59}.

\section{SUMMARY}%
In summary, we have demonstrated that the FMI state of ${\rm
La_{1-x}Sr_{x}MnO_{3}}$ originates from the electron-phonon coupling
induced Peierls instability when the orbital order confines holes to
move along quasi-1D pathways. With this picture, the reentrance of
the FMM state has been well explained quantitatively. The other
experimental features of the FMI sate, such as the temperature
dependents of resistivity and the giant phonon softening, have been
also understood at the qualitative level. Our theory supports the
belief that the intercoupling of hole-orbital-phonon is critical in
understanding the electronic properties of doped magnanites.

\begin{acknowledgments}
Support from the National Natural Science Foundation of China
(Grants No.~10604037) and the National Basic Research Program of
China (Grants No.~2007CB925001) are gratefully acknowledged.
XRW is supported by HK UGC/GRF grants.
\end{acknowledgments}

\end{document}